\documentclass[pdflatex,sn-mathphys-num]{sn-jnl}

\usepackage{graphicx}%
\usepackage{multirow}%
\usepackage{amsmath,amssymb,amsfonts}%
\usepackage{amsthm}%
\usepackage{mathrsfs}%
\usepackage[title]{appendix}%
\usepackage{xcolor}%
\usepackage{textcomp}%
\usepackage{manyfoot}%
\usepackage{booktabs}%
\usepackage{algorithm}%
\usepackage{algorithmicx}%
\usepackage{algpseudocode}%
\usepackage{listings}%
\usepackage{comment}%
\usepackage{lineno}%

\theoremstyle{thmstyleone}%
\newtheorem{theorem}{Theorem}
\newtheorem{proposition}[theorem]{Proposition}%

\theoremstyle{thmstyletwo}%
\newtheorem{example}{Example}%
\newtheorem{remark}{Remark}%

\theoremstyle{thmstylethree}%
\newtheorem{definition}{Definition}%

\begin{document}
\title[\title{Single-Shot Multispectral Mid-Infrared Imaging with Incoherent Light via Adiabatic Upconversion}]{Single-Shot Multispectral Mid-Infrared Imaging with Incoherent Light via Adiabatic Upconversion}


\author[1,2]{\fnm{Daniel} \sur{Beitner}}
\equalcont{These authors contributed equally to this work.}
\author[1]{\fnm{Ziv} \sur{Abelson}}\email{zivabelson@mail.tau.ac.il}
\equalcont{These authors contributed equally to this work.}

\author[2]{\fnm{Eyal} \sur{Hollander}}
\author[1]{\fnm{Omri} \sur{Meron}}
\author[1]{\fnm{Haim} \sur{Suchowski}}\email{haimsu@tauex.tau.ac.il}

\affil[1]{\orgdiv{Condensed Matter Physics Department,
School of Physics and Astronomy}, \orgname{Faculty of Exact Sciences
and Center for Light-Matter Interaction, Tel-Aviv University}, \orgaddress{\city{Tel Aviv}, \postcode{6997801}, \state{State}, \country{Israel}}}

\affil[2]{\orgname{Spiral Photonics}}


\abstract{ Multispectral mid-infrared ($2–5\;\mu m$) imaging is a critical capability across science and technology, offering a window into the vibrational and thermal landscape of matter that is inaccessible to visible sensors. It bridges the microscopic world of molecular interactions with macroscopic sensing technologies, with applications in environmental sensing, defense and molecular diagnostics. However, current mid-IR cameras require cryogenic cooling and exhibit limited pixel resolution, high cost, and restricted spectral access. Optical up-conversion provides a pathway to overcome these limitations, but existing systems typically rely on narrowband phase matching, mechanical scanning, or angular tuning, limiting imaging speed and practicality. Here, we demonstrate the first single-shot, room-temperature multispectral mid-IR imaging of incoherent thermal light enabled by adiabatic sum-frequency conversion. Our system simultaneously converts the entire $2–5\;\mu m$  region into the visible domain, capturing the image on a Silicon detector with spatial resolution below $20\;\mu m$ and high angular tolerance. We validate full-field imaging using a USAF resolution target and demonstrate spectroscopic contrast imaging in dielectric metamaterials by resolving wavelength and polarization dependent scattering resonances, all achieved without scanning, thermal control, or cryogenic operation. This compact and robust approach bridges the gap between laboratory-grade infrared sensors and scalable Silicon-based detection technologies suitable for real-world deployment.}

\keywords{Multispectral Imaging, Upconversion, Infrared Imaging, Adiabatic Poling}



\maketitle

Imaging in the mid-infrared (mid-IR) spectral range is essential for many technologies, spanning environmental monitoring \cite{loder2015focal,reeves2001mid}, spectroscopy \cite{pilling2016fundamental,gabrieli2019near}, medical diagnostics \cite{fernandez2005infrared,kummel2021rapid,walsh2012label}, and remote sensing \cite{habel2024young, ahrer2023early}. This region contains the fundamental vibrational modes of most molecules, offering a uniquely information-rich window for chemical and material analysis \cite{ozaki2021infrared}. Yet, capturing this spectral abundance via compact, robust and uncooled cameras remains technologically challenging. The performance of conventional mid-IR detectors is limited by intrinsic material and electronic constraints. For example, Mercury-Cadmium-Telluride and III–V semiconductor detectors require cryogenic cooling to suppress dark noise \cite{karim2013infrared}, feature relatively large pixel sizes \cite{rogalski2022scaling}, and are costly to fabricate \cite{rogalski2002infrared}, restricting the realization of compact, broadband, and high-resolution infrared cameras suitable for widespread deployment.
\\A promising strategy to overcome these limitations is optical parametric up-conversion \cite{barh2019parametric}, in which mid-IR photons are mixed with a higher-frequency pump field to generate visible or near-infrared photons detectable by standard Silicon sensors. This nonlinear frequency conversion enables room-temperature operation, high spatial resolution, and compatibility with mature Silicon-detector technology. However, broadband up-conversion remains difficult, as strict phase-matching between interacting waves confines efficient conversion to narrow spectral and angular ranges \cite{saleh2019fundamentals}. Approaches that extend this bandwidth, such as temperature tuning \cite{Hu:12, kehlet2015infrared,hogstedt2014low}, angle scanning \cite{Kehlet:15,huot2016upconversion}, or wavelength-swept pumping \cite{demur2018near,tidemand2024broadband}, inevitably introduce additional calibration and tuning requirements, increase system size, and reduce robustness and long-term stability.
\\A robust solution is offered by adiabatically poled nonlinear crystals \cite{suchowski2009robust,suchowski2014adiabatic}, in which the poling period gradually varies along the propagation direction to maintain quasi-phase-matching over a broad frequency and angular range. This approach enables highly efficient and robust frequency conversion without fine tuning, allowing a wide mid-IR bandwidth to be mapped into the visible domain in a single exposure. Previous studies have demonstrated the power and versatility of adiabatic poling using coherent illumination, achieving broadband, multicolor, and time-resolved upconversion imaging with high efficiency and stability \cite{fang2023mid, fang2024wide, huang2022wide, mrejen2020multicolor}. However, despite these successes, the application of adiabatic upconversion utilizing incoherent or broadband thermal illumination, the regime most relevant for practical mid-IR imaging, has remained largely unexplored. 
\\Here, we demonstrate a broadband, single-shot mid-IR imaging system spanning $2-5\;\mu m$ by upconverting incoherent black-body radiation into the visible domain. Our compact, no-moving-parts apparatus combines a Q-switched 1064 nm pump laser with high peak intensity, with an adiabatically poled Lithium-Niobate (APLN) crystal engineered for wide spectral and angular acceptance. This architecture maintains robust phase matching without temperature or angular tuning, enabling the entire mid-IR vibrational band to be captured simultaneously on a standard Silicon detector at room temperature. 
We show, through a series of experiments, near–diffraction-limited spatial resolution ($<20\; \mu m$) across the mid-IR band, while supporting multiband imaging through spectral selection performed in the visible domain. Using this approach, the entire mid-IR bandwidth is converted simultaneously in a single exposure, and specific mid-IR spectral slices are subsequently accessed by filtering the upconverted image in the visible with one-to-one spectral mapping. Moreover, we utilize the apparatus to characterize arrays of subwavelength Silicon-based metamaterials, resolving wavelength and polarization dependent scattering responses that originate from anisotropic resonant optical modes across the mid-IR band. Together, these capabilities establish adiabatic up-conversion as a promising route towards compact, alignment-free, and broadband infrared imaging systems directed at practical applications.

\begin{figure*}[t]
\includegraphics[width=13 cm]{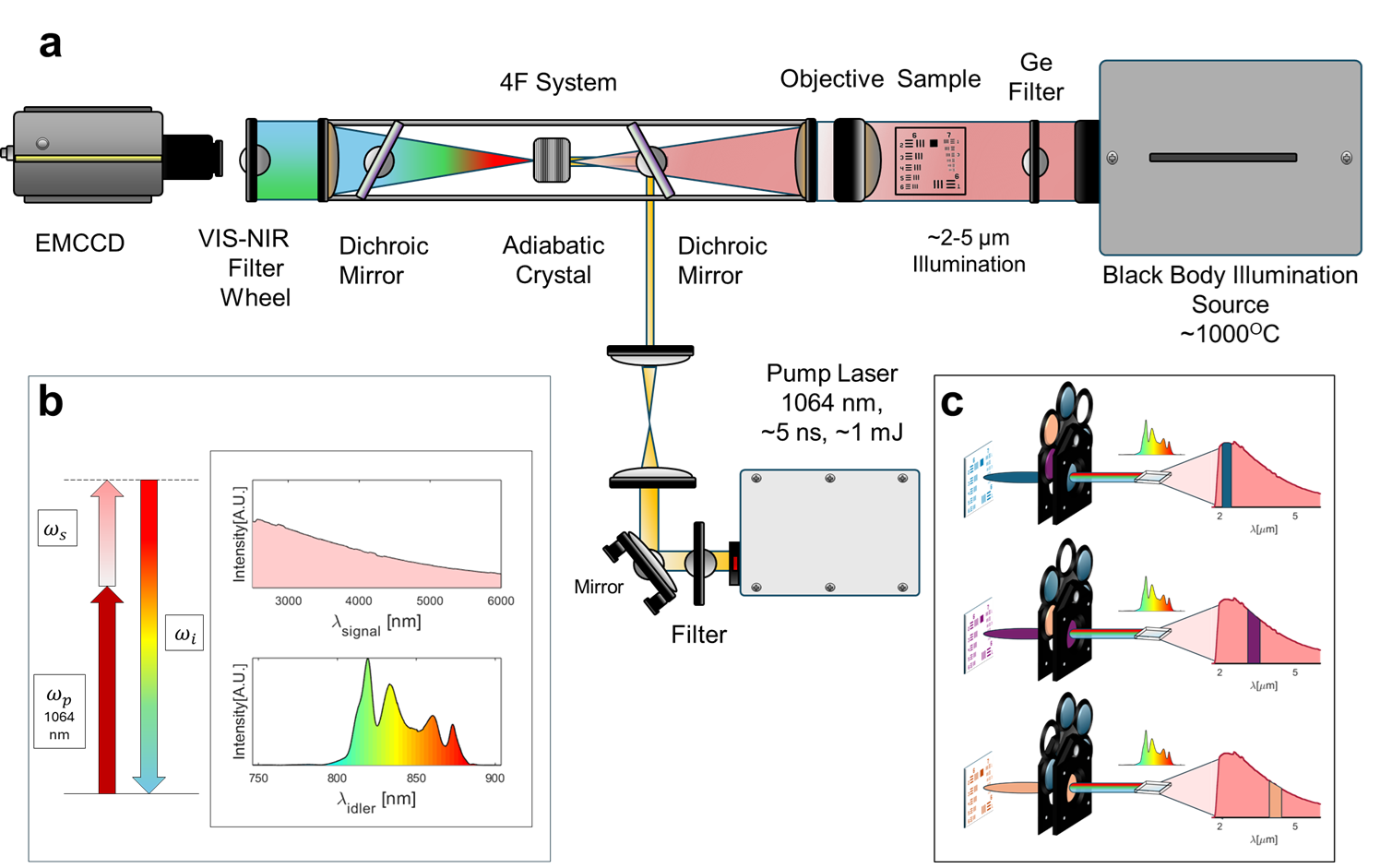}
\caption{\textbf{Broadband adiabatic up-conversion imaging system for incoherent mid-IR illumination.} \textbf{a} Schematic of the single-shot imaging setup. Broadband incoherent (thermal) radiation from a black-body source ($\sim2–5\;\mu m$) illuminates the sample and is relayed through a 4f system. A Q-switched $1064\;nm$ pump beam is combined with the mid-IR image at a dichroic mirror and focused into an adiabatically poled Lithium-Niobate crystal, where sum-frequency generation (SFG) converts the mid-IR spatial information into the visible–near-infrared (VIS–NIR) domain. A second dichroic mirror removes residual mid-IR and pump light, and the up-converted image is recorded on a room-temperature Silicon EMCCD. A VIS–NIR filter wheel enables spectral selection without any mechanical scanning in the mid-IR.
\textbf{b} Energy-conservation diagram showing up-conversion of a broad mid-IR spectrum into a narrower VIS–NIR band through interaction with the $1064\; nm$ pump.
\textbf{c} Conceptual demonstration of multispectral imaging: band-pass filtering in the VIS–NIR reveals different spectral features of the mid-IR scene from a single captured frame.} {\label{Fig1}}
\end{figure*}

\begin{figure}[h]
\centering
\includegraphics[width=1\textwidth]{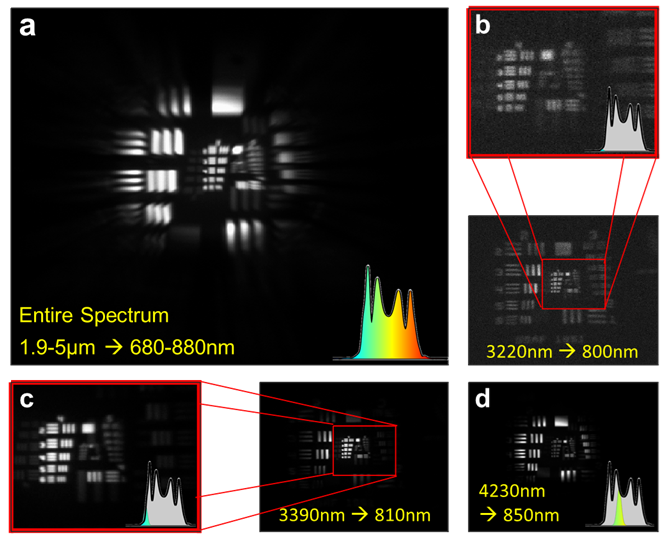}
\caption{\textbf{Multispectral mid-IR imaging of a USAF resolution target using visible-band filtering}.
\textbf{a} Single-shot upconverted mid-IR image of a 1951 USAF resolution test chart, obtained by capturing the entire $\sim 2–5\;\mu m$ band simultaneously. Wavelength-dependent magnification leads to blurring of the finest features in the full-spectrum image.
\textbf{b–d} Narrowband VIS–NIR filtering of the up-converted image isolates different regions of the mid-IR spectrum, improving image quality as chromatic dispersion is reduced. Insets show zoomed views of the smallest resolvable features, along with the corresponding portion of the mid-IR illumination spectrum mapped to the filter passband. Effective mid-IR center wavelengths and bandwidths (FWHM) are indicated below each panel.
These measurements demonstrate near-diffraction-limited ($<20\;\mu m$) resolution over a wide mid-IR spectral range without mechanical or thermal scanning.}\label{Fig2}
\end{figure}

\section*{Results}\label{sec2}
The experimental platform, presented in Fig. \ref{Fig1}, is built to enable broadband mid-IR imaging with a compact, alignment-tolerant architecture. We use a high-peak-power pulsed $1064\; nm$ pump in a small-footprint optical layout, to provide an efficient nonlinear interaction without relying on bulky or actively stabilized laser systems. As shown in Fig. \ref{Fig1}a, broadband mid-IR radiation ($2-5\;\mu m$) from a blackbody source illuminates the object and is relayed by a 4f imaging system into an adiabatically poled Lithium-Niobate crystal, preserving the full spatial information of the scene. 
\\The synchronized $1064\; nm$ pulsed pump beam is spatially overlapped with the mid-IR image inside the crystal, where sum-frequency generation converts the mid-IR image into the visible–near-infrared domain. Residual pump and mid-IR light are rejected by dichroic filtering, and the upconverted image is recorded using a Silicon detector. This approach eliminates the need for mechanical scanning or temperature tuning, while maintaining room-temperature operation with a fully static setup. The resulting low-complexity configuration is robust and practical, yet preserves the wide spectral acceptance and high spatial fidelity that are paramount for the adiabatic conversion concept.
\\The underlying frequency mapping is illustrated in Fig. \ref{Fig1}b, where energy conservation maps the broadband mid-IR spectrum to a corresponding visible band set by the pump wavelength, while the adiabatically varying poling period maintains quasi-phase-matching over a wide spectral and angular range. Consequently, the entire mid-IR bandwidth is converted simultaneously in a single exposure, without angular tuning, temperature control, or scanning. Figure \ref{Fig1}c illustrates how multispectral imaging is realized following the up-conversion stage through spectral selection in the visible domain. A one-to-one nonlinear mapping links each visible wavelength to a corresponding mid-IR wavelength, allowing visible-band filtering to isolate specific mid-IR spectral slices from the same single-shot image while preserving spatial fidelity.
\\Initially, to evaluate system performance, a standard 1951 USAF resolution target was examined using broadband thermal illumination. Figure \ref{Fig2}a shows the single-shot up-converted image obtained when the full $2–5\;\mu m$ mid-IR spectrum is captured simultaneously. In this case, wavelength-dependent magnification \cite{coen2023diffraction}, arising from momentum-conservation-driven dispersion in the up-conversion process , slightly blurs the finest spatial features.
\\Applying narrowband filtering in the VIS–NIR domain isolates selected portions of the mid-IR spectrum and enables observation of spectrally dependent features. As shown in Fig. \ref{Fig2}b-d, this approach limits the wavelength-dependent magnification, leading to a clear improvement in image quality, with features below $20\;\mu m$ clearly resolved, approaching the diffraction limit of our optical design.
\\We note that spectral selection in the upconverted image is fundamentally defined in the frequency domain. In sum-frequency up-conversion with a fixed pump wavelength, the mid-IR and visible bandwidths are equal in frequency ($\Delta \omega _{IR}=\Delta \omega_{upconversion}$).  When expressed in wavelength, this correspondence results in an expanded effective bandwidth in the mid-IR due to the nonlinear relation between frequency and wavelength. Consequently, a $10\;nm$ band-pass filter centered at $\approx 800\; nm$ selects an effective mid-IR bandwidth of approximately $\approx 160\; nm$ around $3.5\; \mu m$, while preserving the intrinsic spatial resolution of the system. 
\\Further reducing the spectral bandwidth could provide additional enhancement of image quality, pushing the boundary towards the intrinsic resolution but also decreases optical throughput, which becomes critical in spectral regions where the up-conversion efficiency is lower. In addition, the field of view varies between each of the spectrally filtered images because of the wavelength-dependent angular acceptance of the adiabatic crystal design (additional examples provided in supplementary information Fig. S1). 

\begin{figure*}[h]
\includegraphics[width=13.1 cm]{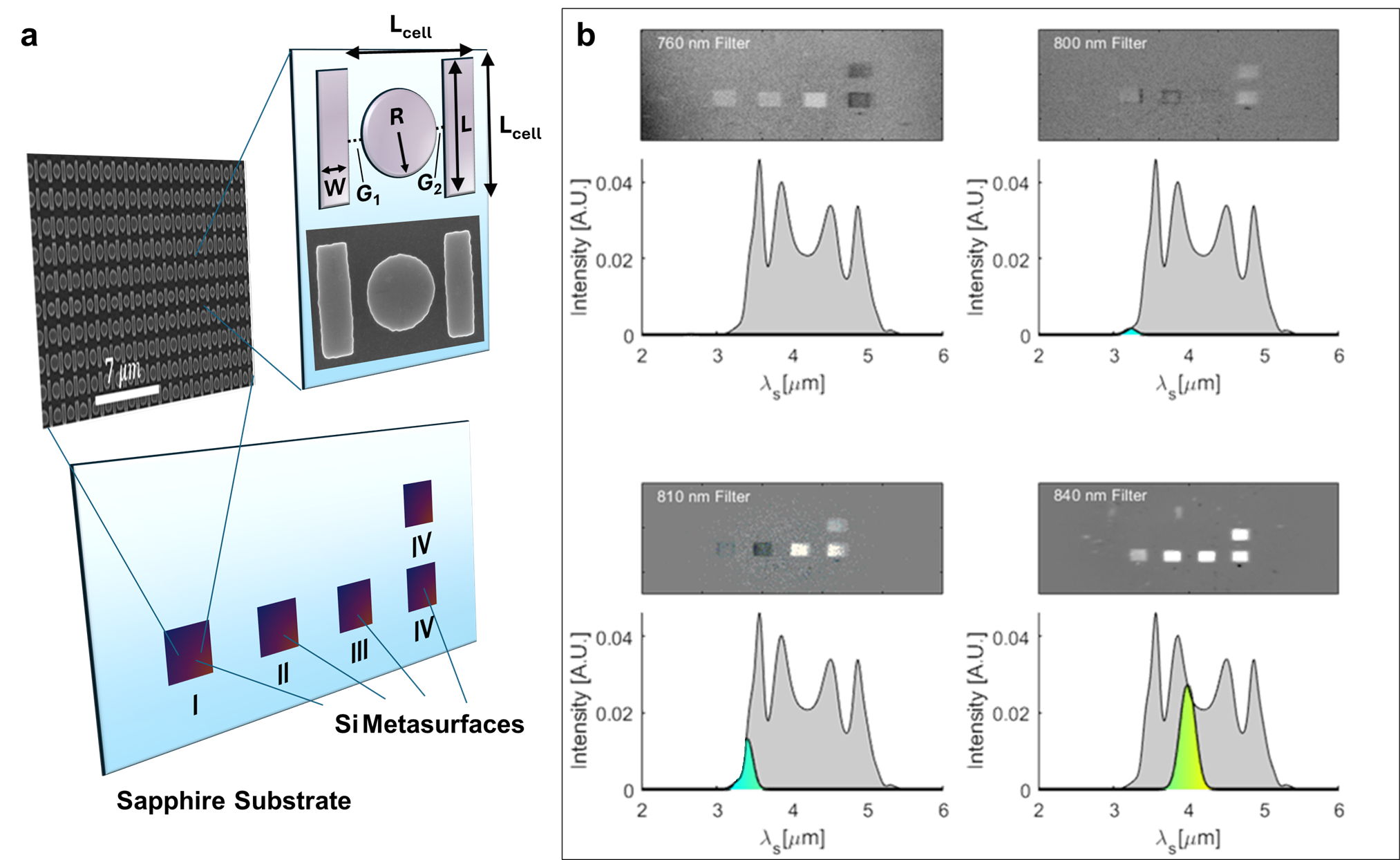}
\caption{\textbf{Multispectral imaging of mid-infrared Mie resonance metamaterials.} \textbf{a} Illustration and characterization of the metamaterial sample. Four different designs of  nanostructure arrays have been fabricated on a Sapphire subsrate. A fifth array, marked with an additional 'IV' label, is a duplicate of the same design. For a single unit cell, the critical dimensions of each of the designs is presented. These dimensions influence the central wavelength and shape of the mid-IR resonance and are elaborated in the Methods section. Additionally, scanning electron microscope images of the nanostructure array and a single unit cell are also provided. 
\textbf{b} Narrowband VIS–NIR filtering of the up-converted image demonstrates a strong scattering response from individual nanostructure arrays, in correlation with their typical resonance peaks. For each of the filtered up-converted images, the corresponding spectral region of the illumination is highlighted. This clearly demonstrates the spectroscopic imaging capabilities of the system.
} {\label{Fig3}}
\end{figure*}
\begin{figure}[h]
\centering
\includegraphics[width=1\textwidth]{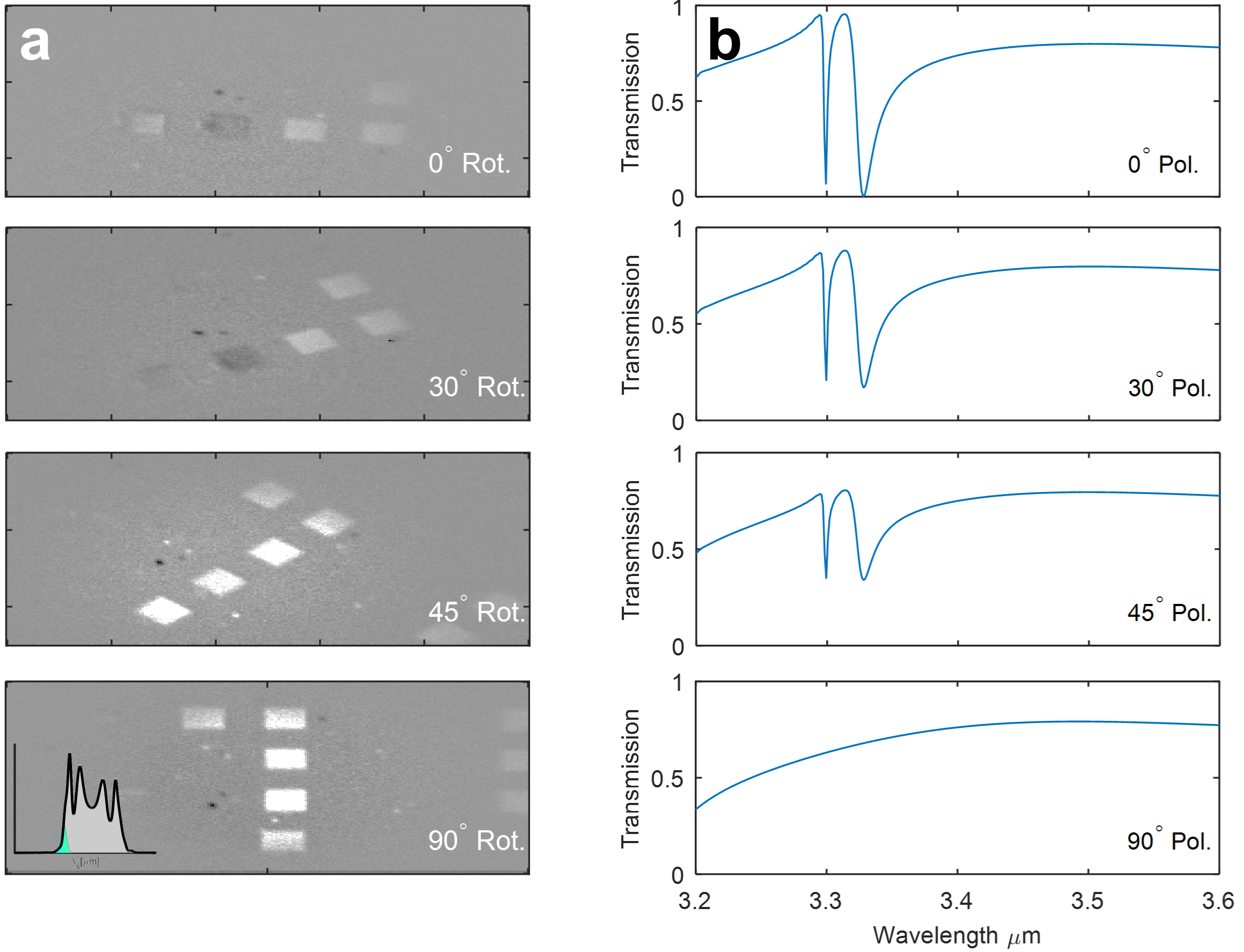}
\caption{\textbf{Polarization-dependent response of mid-infrared Mie-resonant metasurfaces.}
\textbf{a} Spectrally filtered upconverted images of the nanostructure arrays recorded at $810\; nm$ ($10\; nm$ FWHM) for different sample rotation angles, revealing polarization-dependent scattering contrast. The inset indicates the corresponding effective mid-IR spectral range selected by the visible-domain filtering. \textbf{b} Lumerical FDTD simulations of the polarization-resolved spectral response of the all-dielectric nanostructures (design II), showing a resonance that is strongest for a polarization angle of $0^\circ$ and progressively weakens as the polarization is rotated.
}\label{Fig4}
\end{figure}

We extend the validation of the broadband spectral response and utilize the adiabatic up-conversion system to capture and characterize images of a specimen with strong wavelength-dependent optical behavior in the mid-IR. The sample consists of all-dielectric meta-surface arrays \cite{liu2018enhanced,brener2019dielectric} composed of high-refractive-index monocrystalline Silicon nanostructures ($\approx230\;nm$ thick) on a Sapphire ($Al_2O_3$) substrate. By tailoring their lateral dimensions, these nanostructures support high-Q Mie resonances across the full $2–5\;\mu m$ range, originating from the excitation of electric and magnetic multipoles either individually or through modal interference. Four distinct arrays were fabricated, each engineered to resonate at different mid-IR wavelengths, also exhibiting polarization-dependent responses due to geometric asymmetry.
\\Figure \ref{Fig3} shows the upconverted images of four metasurface samples, recorded using $10\; nm$-wide bandpass VIS–NIR filters. Each image isolates a different portion of the mid-IR spectrum after nonlinear mapping and reflects the distinct nanostructure parameters of the individual arrays. Corresponding scanning electron micrographs of the metasurface geometries are also shown. A clear enhancement in scattering from each array is observed only when the selected spectral band overlaps its resonance condition, demonstrating strong agreement with electromagnetic simulations. The ability to identify resonant features even in spectral regions where nonlinear conversion efficiency is reduced highlights the high sensitivity of the adiabatic up-conversion scheme. These results confirm that the platform provides spectral contrast without mechanical scanning, operating as a compact multispectral imager that combines wavelength-resolved visibility with high spatial resolution in a single-shot configuration.
\\Finally, we examine the polarization-dependent response of the metasurface arrays to illustrate the system’s ability to resolve anisotropic resonances in the mid-IR arising from the anisotropic nanostructure geometries. As shown in Fig. \ref{Fig4}a, rotating the sample with respect to the incident polarization produces pronounced variations in the scattering intensity of the individual arrays. These changes directly reflect the symmetry-dependent coupling of the incident field to the multipolar modes supported by the anisotropic nanostructure designs, in good agreement with electromagnetic simulations of the polarization-resolved response (Fig. \ref{Fig4}b). Owing to the polarization-dependent efficiency of the up-conversion process, the imaging system exhibits intrinsic polarization sensitivity. The strong contrast modulation observed across polarization angles demonstrates that the platform can sensitively resolve not only broadband spectral signatures but also directional and symmetry-dependent optical responses.  This capability highlights the versatility of adiabatic up-conversion imaging as a tool for comprehensive material characterization in the mid-IR, simultaneously resolving spatial, spectral, and polarization degrees of freedom without the need for scanning, cryogenic cooling, or specialized spectrometers.

\section*{Discussion}\label{sec3}
The demonstrated architecture establishes adiabatic up-conversion as a practical and high-performance approach for mid-IR imaging, enabling simultaneous spatial, spectral, and polarization contrast in a single exposure. The system operates at room temperature, requires no moving parts or cryogenic cooling, and maintains near-diffraction-limited spatial resolution across the $2-5\; \mu m$ band, confirming the robustness of the wide-acceptance, adiabatically poled crystal design.
\\Beyond the demonstrated performance, the platform provides a flexible and scalable design space that can be adapted to specific application requirements. The adiabatic up-conversion approach is inherently broadband and robust, providing higher conversion efficiency in comparison to existing systems \cite{margules2021ultrafast}, and the nonlinear crystal can be engineered to target different infrared spectral regions through appropriate poling profiles. In this context, the primary system parameters governed by the crystal design are the achievable conversion efficiency and the wavevector acceptance angle, which together define the trade-offs between spectral bandwidth, field of view, and sensitivity. In addition, the operation of the Q-switch pump can be reduced to optimal performance parameters, enabling the implementation of a small footprint device.
\\The use of standard Silicon detectors further enhances the practicality of the approach by leveraging small pixel pitches, large pixel counts, and mature readout technology. This also opens the possibility of implementing pixel-level spectral encoding, analogous to Bayer-filter architectures in visible cameras, in which different spectral bands are assigned to interleaved pixel groups. In such a configuration, multiple mid-IR spectral channels could be recorded simultaneously in a single exposure. \\ In combination with the ability to access multispectral information without mechanical scanning, this enables compact implementations with reduced system complexity and favorable size-weight characteristics. Apart from utilizing all the benefits of Silicon based detectors, the capability of dual use of the same detection technology for a tremendously larger spectral range holds many design and economical advantages. These attributes are particularly relevant for applications in environmental and industrial monitoring, defense and security imaging, and biomedical diagnostics, where passive thermal and emissivity-driven contrasts provide information not accessible at shorter wavelengths. The ability to resolve anisotropic resonant responses in metasurfaces further highlights the platform’s utility for material characterization and functional imaging of nanophotonic structures in the mid-IR and could also transform the field of mid-IR resonant material testing \cite{barho2020heavily,mitchell2024mid}.
\\ Furthermore, the superior performance that this method provides can be utilized to delve deeper into fundamental physical processes residing in the mid-IR. The dynamics of ultrafast thermal mechanisms or mid-IR phonon-polariton electroluminescence, for example, can be probed with greater precision and detail. The ability to also tune the adiabatic process to different regions in the entire IR spectrum provides the flexibility which renders this a comprehensive solution.
\\While the present implementation employs fixed VIS-NIR spectral filtering, the architecture naturally supports refined spectral selectivity through tunable or integrated filtering approaches, as well as compatibility with compact mid-IR illumination sources such as LEDs. More broadly, the flexibility of the adiabatic design allows extension to other infrared bands through appropriate crystal engineering. Together, these features position adiabatic up-conversion as a promising route toward manufacturable, uncooled multispectral and hyperspectral mid-IR imaging systems, bridging the gap between laboratory-grade infrared sensing and widespread real-world deployment.

\bibliography{bibliography}

\end{document}